\documentclass[11pt, onecolumn]{article}

\usepackage{hyperref}
\usepackage{graphicx}
\usepackage{amssymb}
\usepackage{amsmath}

\usepackage{color}
\usepackage{mathrsfs}
\usepackage[refpage, notintoc]{nomencl} 
\usepackage[version=3]{mhchem}
\usepackage{chemarr}

\newcommand{\ddt}[1]{\frac{d #1}{dt}}
\newcommand{\mbf}[1]{\mathbf{#1}}
\newcommand{\R}{\mathbb{R}}

\newcommand{\La}{\mathscr{L}}

\newcommand*{\defeq}{\mathrel{\vcenter{\baselineskip0.5ex \lineskiplimit0pt
                     \hbox{\scriptsize.}\hbox{\scriptsize.}}}%
                     =}

\title{Linear models of activation cascades: analytical solutions and
  coarse-graining of delayed signal transduction}

\author{Mariano
  Beguerisse-D\'iaz$^{1,}$\footnote{Current address: Mathematical Institute, University of Oxford, Oxford OX2 6GG, beguerisse@maths.ox.ac.uk} ,
  Radhika Desikan\,$^{2}$ and Mauricio
  Barahona\,$^{1,}$\footnote{m.barahona@imperial.ac.uk}\\ $^{1}$Department
  of Mathematics, Imperial College London, London SW7 2AZ, United
  Kingdom\\ $^{2}$Department of Life Sciences, Imperial College
  London, London SW7 2AZ, United Kingdom}

\date{}

\begin{document}

\maketitle

\begin{abstract}
  Cellular signal transduction usually involves activation cascades, the sequential activation of a series of proteins following the reception of an input signal. Here we study the classic model of weakly activated cascades and obtain analytical solutions for a variety of inputs. We show that in the special but important case of optimal-gain cascades (i.e., when the deactivation rates are identical) the downstream output of the cascade can be represented exactly as a lumped nonlinear module containing an incomplete gamma function with real parameters that depend on the rates and length of the cascade, as well as parameters of the input signal.  The expressions obtained can be applied to the non-identical case when the deactivation rates are random to capture the variability in the cascade outputs.  We also show that cascades can be rearranged so that blocks with similar rates can be lumped and represented through our nonlinear modules.  Our results can be used both to represent cascades in computational models of differential equations and to fit data efficiently, by reducing the number of equations and parameters involved.  In particular, the length of the cascade appears as a real-valued parameter and can thus be fitted in the same manner as Hill coefficients.  Finally, we show how the obtained nonlinear modules can be used instead of delay differential equations to model delays in signal transduction.
  \end{abstract}

\section{Introduction}

\begin{figure}[t]
  \centerline{\includegraphics[height=0.4\textwidth]{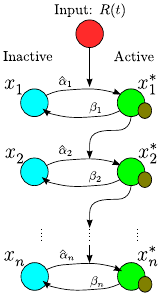}}
    \caption{{\bf A typical protein activation cascade of length $n$.}
      The proteins (nodes) in the cascade can either be in an inactive
      ($x_i$) or active ($x^*_i$) state.  An external signal $R(t)$
      activates the first node. Once a node is active, it activates
      the next component in the cascade until the end. The activation
      rate of each $x_i$ is $\alpha_i$, and the deactivation rate of
      each $x^*_i$ is $\beta_i$. Image adapted
      from~\cite{Heinrich2002}.}
    \label{fig:MAPKcascade}
\end{figure}

Activation cascades are pervasive in cellular signal transduction
systems~\cite{Heinrich2002, Marks2009}. In its simplest form, an
activation cascade comprises a set of components (typically proteins)
that become sequentially activated in response to an external stimulus
(Fig.~\ref{fig:MAPKcascade}). These systems have been the subject of
numerous studies, experimental and theoretical~\cite{Chang2001,
  Chaves2004, Feliu2011, Heinrich2002, Huang1996, Kholodenko2000,
  Tyson2003, Zhang2001a}.  The role of activation cascades in cellular
signal transduction is manifold.  Cascades can relay, amplify, dampen
or modulate signals in order to achieve a variety of cellular
responses.  One of the best studied examples of such a system is the
{\it mitogen-activated protein kinase} (MAPK) cascade, which plays a
central role in key cellular functions, such as regulation of the cell
cycle, stress responses and apoptosis~\cite{Marks2009}.

Models of activation cascades are known to exhibit a range of
nonlinear behaviours, including ultrasensitivity~\cite{Huang1996,
  Li2010} and multistability~\cite{Feliu2011, Thomson2009}.
Linearised models of cascades~\cite{Heinrich2002} (the so-called
`weakly-activated' regime studied here) are also of theoretical
interest, and have been studied to evaluate signalling
times~\cite{Mazza2014}, signal specificity~\cite{Bardwell2007} and
optimal gain~\cite{Chaves2004}.  Such linearised descriptions of
cascades often appear as part of larger and more complicated models,
and have been shown to be useful in model-reduction
techniques~\cite{Herath2015}.  Hence obtaining coarse-grained
representations of such cascades would be useful not only to simplify
their mathematical analysis but also computationally, to allow for
compact implementations in models for Systems Biology.  Furthermore,
weakly-activated cascades are of importance in quantitative biology as
they have been observed experimentally~\cite{Munshi2013}.  In this
context, it would be desirable to estimate the length of an unobserved
cascade from data without having to create and fit several models,
each with a different number of equations to represent the varying
length of the cascade.

Here we present a study of analytical solutions of ordinary
differential equation (ODE) models of linear activation cascades.
First, we obtain general solutions for weakly-activated cascades. We
then focus on the case when the gain of the cascade is optimal (i.e.,
when all deactivation rates are identical), and find that a lower
incomplete gamma function with only three real-valued parameters
represents the output of the entire cascade.  We exemplify the use of
this coarse-grained solution to describe the downstream output induced
by several time-dependent inputs of interest, including step
functions, exponentially decaying signals, Gaussian inputs and
periodic stimuli.  We also show that the obtained solution has
real-valued parameters directly linked to the length and filtering
properties of the cascade, and can thus be used to fit data capturing
efficiently the delay and distortion introduced by the cascade.  We
also explore the application of our results to non-optimal cascades,
i.e., when the requirement of identical deactivation rates is relaxed.
When only one deactivation rate is different, the equations can be
reordered, so that a lumped gamma function representation can be used
for the block of identical proteins without altering the final output
of the cascade. We also show that when the deactivation rates are
randomly distributed, the gamma function can still be used to
represent the distribution of the outputs of the cascade.  Finally, we
show how the gamma function representation of a cascade can be used as
a computationally efficient replacement of delay differential
equations.
 
\section{Weakly activated cascades and their gamma function solution}

Consider a cascade involving $n$ components that are activated in
succession.  Upon perception of the input signal $\hat{R}(t)$, the
first inactive component ($x_1$) is transformed into its activated
form ($x^*_1$), which then activates the next component
($x_2$). Sequential activation of $x_i$ by $x^*_{i-1}$ continues until
the end of the cascade. The output of the cascade of length $n$ is the
activated form of the last component, $x^*_n$.  In the case of the
MAPK cascade, the components are proteins, and the activation
corresponds to a post-translational modification, i.e.,
phosphorylation. However, the formalism can also describe other
sequential biochemical processes with similar functional
relationships, e.g., $n$-step deoxyribonucleic acid (DNA)
unwinding~\cite{Lucius2003}.

If we use mass-action kinetics without an intermediate complex to
describe protein activation, the reaction describing the activation of
$x_1$ is
\begin{equation*} 
  R + x_1 \ce{->[\hat{\alpha}_1]} x^*_1 + R,
\end{equation*}
and for the rest of the proteins $x_i$ ($i=2,\hdots, n$) we have
\begin{equation*}
  x^*_{i-1} + x_i \ce{->[\hat{\alpha}_i]} x^*_{i-1} + x^*_{i}.
\end{equation*}
We also assume that all proteins deactivate spontaneously with constant
rate:
\begin{equation*}
    x^*_i \ce{->[\beta_i]} x_{i}.
\end{equation*}
 The system of nonlinear ODEs describing the time evolution
of the full activation cascade is~\cite{Heinrich2002}:
\begin{align}
  \ddt{x^*_1} &= \hat{\alpha}_1 R(t)(T_1 - x^*_1) - \beta_1x^*_1,
  \nonumber \\ \ddt{x^*_2} &= \hat{\alpha}_2 x^*_1(T_2 - x^*_2) -
  \beta_2x^*_2, \label{eq:nonlinear} \\ & \vdots \nonumber
  \\ \ddt{x^*_n} &= \hat{\alpha}_n x^*_{n-1}(T_n - x^*_n) -
  \beta_nx^*_n, \nonumber
\end{align}
where we have defined the total amount of each protein
$T_i=x_i+x_i^*$, so that the inactive form is $x_i=(T_i - x^*_i)$.  We
also assume that the model operates over time scales where there is no
significant protein production, so that the amount of each protein
$T_i$ can be considered constant.  If the timescales are such that the
total amount of protein varies significantly, then each $T_i$ would
have to be described by its own ODE according to additional biological
knowledge.

\paragraph{The general solution for weakly activated cascades.}
As shown in Ref.~\cite{Heinrich2002}, in the {\it weakly activated}
regime $T_i \gg x_i^*$, one takes the approximation $(T_i - x_i^*)
\approx T_i$, and the original system~(\ref{eq:nonlinear}) can be
rewritten as a driven linear system:
\begin{equation}
  \ddt{\mbf{x^*}} = \mbf{A x^*} + \alpha_1R(t) \,\mbf{e}_1,
  \label{eq:weak-ac-syst}
\end{equation}
where $\mbf{x^*} = [x^*_1, \dots, x^*_n]^T$,  
$\mbf{e}_1 = [1,0,\dots,0]^T$ is the first $n \times 1$ vector of
the canonical basis, and
the $n \times n$ rate matrix
$\mbf{A}$ is:
\begin{equation}
  \mbf{A} = \left[
  \begin{array}{cccc}
    -\beta_1 &  &  &  \\
    \alpha_2 & -\beta_2 &  &  \\
    & \ddots & \ddots &  \\
    &  & \alpha_n & -\beta_n  
  \end{array} \right],
  \label{eq:A-matrix}
\end{equation}
where $\alpha_i = \hat{\alpha_i}T_i, \, \forall i$.

This system can be solved using the Laplace transform with auxiliary
variable $s$.  If the cascade receives an integrable input $R(t)$, it
is easy to show that the Laplace transform of the $k^{\mathrm{th}}$
protein is
\begin{equation}
  \La(x_k^*) =
  \overbrace{\frac{\alpha_{(k)}^k\La(R)}{\prod_{i=1}^{k}(\beta_i +
      s)}}^{\text{Dynamics from rest}} +
  \overbrace{\sum_{i=1}^k\frac{\alpha_{(k)}^k}{\alpha_{(i)}^i}\frac{x_i^*(0)}{\prod_{j=i}^k(\beta_j
      + s)}}^{\text{Correction for initial condition}}.
  \label{eq:xn-lap-trans}
\end{equation}
The first term on the right-hand side corresponds to the Laplace
transform of $x_{k}^*(t)$ for initial conditions
$x_i^*(0)=0,\,\forall i\leq k$ (i.e., the cascade starts from rest),
and the second term contains the correction for non-zero initial
conditions.  The term $\alpha_{(k)}$ is the geometric mean of the
activation rates up to $k$:
\begin{equation}
  \alpha_{(k)} = \left(\prod_{j=1}^k\alpha_j\right)^{1/k}.
  \label{eq:alpha-k}
\end{equation}
Note that if $\beta_i\neq \beta_j, \, \forall \, i,j$ then
\begin{equation*}
  \prod_{j=1}^k(\beta_j+s)^{-1} 
  = \sum_{j=1}^k\frac{\beta_{(-j)}^{(k)}}{\beta_j + s},
\end{equation*}
where
\begin{equation*}
  \beta_{(-j)}^{(k)} = \prod_{\substack{i=1 \\ i\neq
      j}}^k(\beta_j-\beta_i)^{-1} \in \R, \quad \beta_{(-j)}^{(0)} \defeq 1,
\end{equation*}
is a constant that depends only on the deactivation rates.
Now we can express equation~(\ref{eq:xn-lap-trans}) as 
\begin{equation}
  \La(x_k^*)= \sum_{i=1}^k \left\{
  \frac{\alpha_{(k)}^k\beta_{(-i)}^{(k)}\La(R)}{\beta_i + s} +
  \frac{\alpha_{(k)}^k}{\alpha_{(i)}^i} \sum_{j=i}^k
  \frac{\beta_{(-j)}^{(k)}}{\beta_{(-j)}^{(i-1)}}
  \frac{x_i^*(0)}{\beta_j +s} \right\}.
  \label{eq:lap-xn}
\end{equation}
Using linearity and the convolution properties of the Laplace
transform, the output of the cascade is finally obtained as:
\begin{equation}
  x_n^*(t) = \alpha_{(n)}^n\sum_{i=1}^n \left\{ \beta_{(-i)}^{(n)} 
  \left(R \ast e^{-\beta_i t}\right)(t) +
  \frac{1}{\alpha_{(i)}^i} \sum_{j=i}^k 
  \frac{\beta_{(-j)}^{(k)}}{\beta_{(-j)}^{(i-1)}} x_i^*(0)e^{-\beta_j t}
  \right\}
  \label{eq:xn-gen-sol}
\end{equation}
where
\begin{equation*}
  \left(R\ast e^{-\beta_i t}\right)(t) =
  \int_0^{t}e^{-\beta_i(t-\tau)}R(\tau)\mathrm{d}\tau =
  \int_0^{t}e^{-\beta_i \tau}R(t -\tau)\mathrm{d}\tau,
\end{equation*}
and the
pre-factor incorporates the product of all the activation rates,
\begin{equation*}
\alpha_{(n)}^n = \prod_{i=1}^n \alpha_i.
\end{equation*}

Although Eq.~\eqref{eq:xn-gen-sol} describes the evolution of a
general initial condition, in this study we will assume henceforth
that the cascade is initially fully inactive (i.e., $x_i^*(0)=0, \,
\forall i$). In the cases when $x_i^*(0)\neq 0$, then the exponential
correction introduced by the initial conditions can be incorporated to
the calculations.

\paragraph*{\textit{Example:}} If a linear cascade is subject to a
constant stimulus given by the step function $R(t)=1, \, t \geq
0$, and $x_i^*(0)=0 \,\, \forall \, i$, Eq.~\eqref{eq:xn-gen-sol}
shows that the output of the last protein in the cascade is given by:
\begin{equation}
    x_n^*(t)  = \alpha_{(n)}^n\sum_{i=1}^n\frac{\beta_{(-i)}^{(n)}}{\beta_i}
    \left[1 - e^{-\beta_i t}\right].
    \label{eq:xn-gen-sol-const-stim}
\end{equation}

\paragraph{Optimal linear cascades}
Activation cascades are substantial modules of the cell-signalling
machinery and, as such, they should be efficient in minimising the use
of energetic resources, such as adenosine triphosphate (ATP), or of
cellular building blocks, such as amino acids.  In
Ref.~\cite{Chaves2004} it was shown that when a weakly activated
cascade~(\ref{eq:weak-ac-syst}) is required to provide a given gain,
the amplification is achieved optimally when the number of steps in
the cascade (e.g., the number of proteins) is finite and all
deactivation rates are equal, i.e., $\beta_i=\beta,\, \forall i$.
This result means that arbitrarily long cascades are not useful for
cells when a particular amplification gain from external signals is
required.  For an optimal cascade, the rate matrix in
Eq.~(\ref{eq:weak-ac-syst}) becomes
\begin{equation}
  \mbf{\tilde A} = \left[
  \begin{array}{cccc}
    -\beta &  &  &  \\
    \alpha_2 & -\beta &  &  \\
    & \ddots & \ddots &  \\
    &  & \alpha_n & -\beta  
  \end{array} \right].
  \label{eq:A_constant-matrix}
\end{equation}

\section{Linear cascades under different input functions}

We now consider the time-dependent output of a cascade under
four different inputs of biological interest.

\subsection{Step-function stimulus}

In an experimental setting, one often studies the response of a
biological system to a step-function stimulus such as constant
temperature, light or treatment started at time $t=0$. In this case,
the stimulus is:
$$R(t) = 1, \, t \geq 0,$$  
and the solution to~(\ref{eq:weak-ac-syst}) with initial condition
$\mbf{x^*}(0)=\mbf{0}$ is:
\begin{equation}
   \mbf{x^*}(t) = 
  \alpha_1 \mbf{A}^{-1} \left[e^{t\mbf{A}} -  \mbf{I}_n \right]\mbf{e}_1,
  \label{eq:weak-const-sol}
\end{equation}
where $\mbf{I}_n$ is the $n\times n$ identity matrix, and $e^{t\mbf{
    A}}$ is the matrix exponential.  
    
If the cascade is optimal (i.e., $\mbf{A}=\tilde{\mbf{A}}$), the
Laplace transform of the last protein given by~(\ref{eq:xn-lap-trans})
becomes 
\begin{equation*}
  \La(x_n^*) = \frac{\alpha_{(n)}^n}{s(s + \beta)^n},
\end{equation*}
and taking the inverse transform we get:
\begin{equation}
  x_n^*(t) = \left(\frac{\alpha_{(n)}}{\beta}\right)^n \mathrm{P}(n,
  \beta t),
  \label{eq:weak-xn-const-sol}
\end{equation}
where 
\begin{equation}
\mathrm{P}(n, \beta t) = \left(1 - e^{-\beta
  t}\sum_{k=0}^{n-1}\frac{(\beta t)^k}{k!}\right)
  \label{eq:inc-gamm-func-n}
\end{equation}
is the normalised lower incomplete gamma function whose general form
is~\cite{Paris2010}:
\begin{equation}
  \mathrm{P}(a,t) = \frac{\gamma(a,t)}{\Gamma(a)},
   \label{eq:norm-inc-gamma-func}
\end{equation}
where $\Gamma(a)$ is the gamma function
and
\begin{equation*}
  \gamma(a, t) = \int_0^t e^{-s}s^{a-1} \mathrm{d}s, \qquad
  \mathrm{Re}(a)>0.
\end{equation*}

\subsection{Exponentially decreasing stimulus}

When the first protein in the cascade is subject to an exponentially
decaying stimulus (e.g., when the input is a reactive molecule or a
molecule that becomes metabolised, or if the receptors become
desensitised)
$$R(t) =  e^{-\lambda t}, \,\, t \geq 0,$$ then the solution
to~(\ref{eq:weak-ac-syst}) with initial condition
$\mbf{x^*}(0)=\mbf{0}$ is
\begin{equation}
  \mbf{x^*}(t) = \alpha_1\left[e^{t\mbf{A}}- e^{-\lambda
      t}\mbf{I}_n\right] \mbf{A}^{-1} \left[\mbf{I}_n +
    \lambda\mbf{A}^{-1} \right]^{-1}\mbf{e}_1.
  \label{eq:weak-ac-syst-expstim-sol}
\end{equation}

If we assume that the cascade is optimal ($\mbf{A} = \mbf{\tilde
  A}$), then
\begin{equation*}
  \La(x_n^*) = \frac{\alpha_{(n)}}{(s+\lambda)(s+\beta)^n}
\end{equation*}
and the output of the cascade is given by:
\begin{equation}
  x_n^*(t) = \left\{
  \begin{array}{ll}
    \left(\frac{\alpha_{(n)}}{\beta-\lambda}\right)^n \, e^{-\lambda t}
    \, \mathrm{P}(n, (\beta -\lambda)t) & \mathrm{if}~\beta \neq \lambda  \\
    \\
  \frac{1}{\Gamma(n+1)} \left(\alpha_{(n)} t \right)^n \,  e^{-\beta t} & 
  \mathrm{if}~\beta=\lambda ,
  \end{array} \right.
  \label{eq:weak-xn-expdecay-sol}
\end{equation}
where $\alpha_{(n)}$ is defined in~(\ref{eq:alpha-k}).  As in the case
of constant stimulus, the solution is also given in terms of the lower
incomplete gamma function.

\subsection{Periodic stimulus}

In certain experimental settings, we are interested in the response of
a system to a periodic stimulus, e.g., circadian rhythms or day/night
cycles~\cite{Locke2005}.  Let us consider a linear cascade of length
$n$ with periodic input
$$R(t) = 1+\sin(\omega t),$$ which oscillates
between 0 and $2$ with mean $1$ and frequency $\omega >
0$.  From a resting initial condition, the solution to
Eq.~(\ref{eq:weak-ac-syst}) is:
\begin{equation}
  \mbf{x^*}(t) = \alpha_1\mbf{V}^{-1}\left[
    \left(e^{t\mbf{A}}-\mbf{I}_n\right)\mbf{V} -\left(\sin{(\omega
      t)}\mbf{I}_n + \omega\cos{(\omega t)}\mbf{A}^{-1}\right) +
    \omega\mbf{A}^{-1}e^{t\mbf{A}}\right] \mbf{A}^{-1}\mbf{e}_1,
  \label{eq:weak-ac-sin-stim-sol}
\end{equation}
where $\mbf{V} = \left(\mbf{I}_n + \omega^2\mbf{A}^{-2}\right)$.  

When the cascade is optimal ($\mbf{A}=\mbf{\tilde{A}}$), the explicit
solution for the $n$-th protein in the cascade is:
\begin{align}
  x_n^*(t) 
  & = \left(\frac{\alpha_{(n)}}{\beta}\right)^n 
  \left[\mathrm{P}(n,
  \beta t) + \left( \frac{\beta}{r}\right)^n \left( \sin
    \left(\omega t - n\theta \right) - e^{-\beta t} \sum_{k=0}^n
    \frac{(tr)^k}{k!} T_{n+k}(\cos\theta)\right)\right], 
  \label{eq:weak-xn-sin-sol1} 
\end{align}
where $r=\sqrt{\beta^2 + \omega^2}$, $\theta =
\arctan{\left(\beta/\omega\right)}$, and  the $T_{n+k}(\cos\theta)$ are
the Chebyshev polynomials evaluated at~$\cos\theta$.  

Asymptotic limits provide useful insights.  
When the frequency is large compared to the deactivation rate, i.e.,
$\omega \gg \beta$, then $\beta/r \simeq 0$, $\theta \simeq 0$ and
we obtain:
\begin{equation*}
\text{if $\omega \gg\beta$,} \quad {x}_n^*(t) \simeq
\left(\frac{\alpha_{(n)}}{\beta}\right)^n \mathrm{P}(n,
  \beta t).
\end{equation*}
Hence for large frequencies the oscillations in
Eq.~\eqref{eq:weak-xn-sin-sol1} are filtered out, and the solution
approaches the response to the step function given by
Eq.~\eqref{eq:weak-xn-const-sol}.  Conversely, when the deactivation
of the proteins dominates the frequency (i.e., $\beta \gg \omega$) the
behaviour of ${x}^*_n$ will be dominated by the sinusoidal input.

In general, asymptotically as $t \to \infty$, the cascade acts broadly as
a filter with an overall amplification
$\left(\alpha_{(n)}/\beta\right)^n$, and an oscillatory term 
attenuated by a factor $(\beta/r)^n$ with a delay phase $n \theta$:
\begin{equation}
\text{as $t \to \infty$,} \quad
  {x}_n^*(t) = \left(\frac{\alpha_{(n)}}{\beta}\right)^n \left[ 1
    + \left(\frac{\beta}{r}\right)^n\sin{(\omega t - n\theta)}\right],
  \label{eq:weak-xn-sin-sol-longterm}
\end{equation}
where we have used the fact that $\lim_{t\to\infty} \mathrm{P}(n,\beta t) =1$.  
Note that $\beta/r = 1/\sqrt{1 + (\omega/\beta)^2} < 1$, which implies
that ${x}^*_n(t)>0$ for all $t$. 
Cascades with more complicated temporal stimuli can be analysed
similarly using the Fourier series expansion of $R(t)$.

\subsection{Gaussian stimulus}

Gaussian input functions are employed to represent drug intake and
other such signals.  Consider a cascade of length $n$ with input
\begin{equation}
  R(t) =  e^{-\zeta(t-\mu)^2},
  \label{eq:gaussian}
\end{equation}
which describes a bell-curve centred at $t=\mu$, with height $1$ and
amplitude $\zeta$. The solution of Eq.~(\ref{eq:weak-ac-syst}) from
inactive initial conditions under this input is then given by:
\begin{equation}
  \mbf{x}^*(t) = \alpha_1e^{(t-\mu)\mbf{A}}\left\{
    \sum_{k=0}^\infty\frac{(-1)^k}{k!}\sum_{h=0}^k{k \choose h}\left(
      (t-\mu)^{2k-h+1} - (-\mu)^{2k-h+1} \right)
    \frac{\zeta^{k-h}}{2k-h+1}\mbf{A}^h\right\}\mbf{e}_1.
\end{equation}

When a Gaussian input $(2 \pi \sigma^2)^{-1/2} e^{-\frac{(t-\mu)^2}{2
    \sigma^2}}$ becomes increasingly narrow (i.e., $\sigma \to 0$), it
approaches in the limit a Dirac delta function:
$R(t)=\delta(t-\mu)$. In that case, from Eq.~\eqref{eq:xn-gen-sol}
the solution for the $n$-th protein is:
\begin{equation}
  x_n^*(t) = \left\{ \begin{array}{ll}
      0 & t < \mu, \\
      \alpha_{(n)}^n \sum_{i=1}^n\beta_{(-i)}^{(n)}e^{-\beta_i(t- \mu)} & t\geq \mu
      \end{array} \right.
\end{equation}

\section{Applications of the analytical solutions to the
  coarse-grained modelling of cascades}

\subsection{Model simplification and parameter fitting}
\label{sec:param-reduction}

\begin{figure}[tp]
  \centerline{\includegraphics[width=0.9\textwidth]{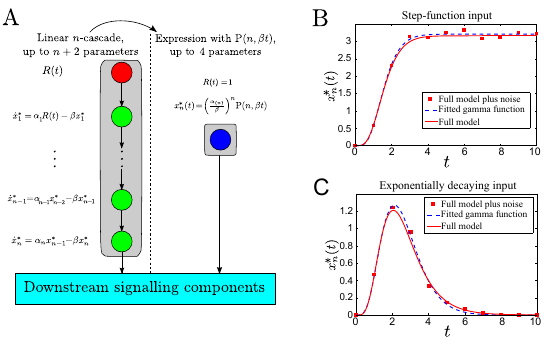}}
    \caption{{\bf Simplification of a linear activation cascade and
        fitting with incomplete gamma functions.}  {\bf A}:~Schematic
      of an optimal linear cascade~\eqref{eq:weak-ac-syst} and its
      corresponding equivalent output
      function~\eqref{eq:weak-xn-const-sol} under a step-function
      input.  The output of the cascade, $x_n$, relays the signal to
      downstream components of the pathway.  Whereas the full
      $n$-dimensional model of the cascade has up to $n+2$ parameters
      ($\alpha_i$, $\beta$, $n$), the condensed expression for the
      output has three parameters $\alpha_{(n)}$, $\beta$, $n$.
      Fitting time-courses of a cascade with two different inputs:
      \textbf{B} a step function and \textbf{C} an exponentially
      decaying stimulus.  In both cases, we considered an optimal
      cascade with $n=5$ components and parameters $\alpha_1=3$,
      $\alpha_i=4$ for $i=2,\dots,5$, and $\beta=3$.  The
      step-function input was $R(t)=1, t \geq 0$ and the exponentially
      decaying input was $R(t)=e^{-\lambda t}$ with $\lambda =1$.  The
      red continuous lines indicate the solutions to the full system
      of $n$ ODEs.  The red squares are `noisy data' generated from
      the full model: $x_5(t)$ sampled at $t=\{0,1,\dots,10\}$ with
      additive Gaussian noise with standard deviation $\sigma =0.05$.
      The blue dashed lines are fits of the noisy data using the
      corresponding incomplete gamma function expressions,
      Eqs.~\eqref{eq:weak-xn-const-sol}~and~\eqref{eq:weak-xn-expdecay-sol}.
      The fits were carried out using the Squeeze-and-Breathe
      algorithm~\cite{Beguerisse2012b}.  }
    \label{fig:networkReplace}
\end{figure}

The expressions of the cascade output, $x_n^*(t)$, obtained in the
previous sections can be used to fit activation data to a small number
of parameters.  Rather than fitting observations to an entire module
of ODEs with $n \in \mathbb{N}$ components, the expressions with the
gamma function contain three parameters ($\alpha_{(n)}$, $\beta$, $n$)
to describe an optimal cascade, and possibly other real parameters
associated with the input (e.g., $\lambda$ for the exponentially
decaying input, or $\omega$ for the periodic stimulus).  In
particular, note that the first argument of the incomplete gamma
function~(\ref{eq:norm-inc-gamma-func}), which is linked to the
cascade length, is a positive \emph{real
  number}~\cite{Abramowitz/Stegun}.  Hence when fitting data (see
Fig.~\ref{fig:networkReplace}), the estimated length of the cascade is
turned into a real-valued parameter $\breve{n} \in \mathbb{R}$,
similarly to what is done with Hill coefficients to represent multiple
mechanistic steps~\cite{Cornish-Bowden2004}.

In Figure~\ref{fig:networkReplace}, we present the application of this
approach to the fitting of the output of an optimal cascade with two
different inputs.  We start by generating simulated data from a
cascade of length $n=5$ with parameters $\alpha_1=3$, $\alpha_i=4$ for
$i=2,\dots,5$ (so that $\alpha_{(n)}=3.776$), and $\beta_i=\beta=3$
for $i=1,\dots,5$.  One cascade is subject to a constant stimulus
$R(t)=1$ and the other to an exponentially decaying input
$R(t)= e^{-\lambda t}$ with $\lambda =1$.  We solve numerically the
$n$-dimensional system of equations~(\ref{eq:weak-ac-syst}) for both
inputs (continuous lines in Fig.~\ref{fig:networkReplace}B,C), and
then we generate `observations' by sampling the output $x_5(t)$ at
times $t=\{0,1,\dots,10\}$ with additive Gaussian noise drawn from
$\mathcal{N}(0,0.05^2)$.  We consider these samples as our `noisy
data' (squares in Fig.~\ref{fig:networkReplace}B,C) and we fit the
gamma function expressions~(\ref{eq:weak-xn-const-sol})\footnote{We
  used the Matlab command {\tt gammainc} to evaluate the lower
  incomplete gamma function.} and~(\ref{eq:weak-xn-expdecay-sol}),
respectively, using a Matlab implementation of the Squeeze and Breathe
evolutionary Monte-Carlo method which is especially appropriate for
time-course series~\cite{Beguerisse2012b}\footnote{Code available
  from~\url{http://people.maths.ox.ac.uk/beguerisse/}}.  The dashed
lines in Fig.~\ref{fig:networkReplace}B show the fits to both cascade
outputs, and the estimated values are close to the `true' ones: for
the constant stimulus cascade, the fitted values are
$\breve{\alpha}_{(n)}\approx 4.068$, $\breve{\beta} \approx 3.281$,
and $\breve{n} \approx 5.418$; for the exponentially decaying
stimulus, the estimated values are
$\breve{\alpha}_{(n)} \approx 3.317$, $\breve{\beta} \approx 2.177$,
$\breve{n} \approx 4.600$, and $\breve{\lambda} \approx 2.177$.

\subsection{Application to near-optimal cascades with random deactivation rates}

Strict optimality of cascades~\cite{Chaves2004} requires that all
deactivation rates of the proteins be identical (i.e., $\beta_i =
\beta$ for all $i$).  Likewise, our expression for the cascade output
in terms of the incomplete gamma function is only strictly valid under
the same assumption.  Naturally, it is unreasonable to expect
identical rates in a biological system. Therefore we ask the question:
if we relax this condition and allow each $\beta_i$ to be an
independent and identically distributed (iid) random variable with
mean $\bar{\beta}$, can we still approximate the output of the module
with a gamma function?

\begin{figure}[t]
  \centerline{\includegraphics[width=0.4\textwidth]{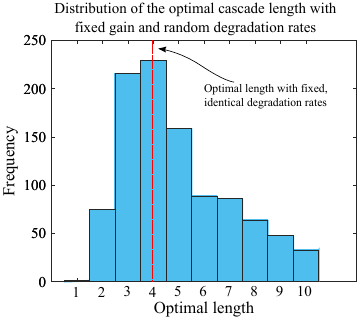}}
    \caption{{\bf Distribution of optimal cascade lengths for
        non-identical (random) deactivation rates.} Simulation of 1000 random
      sets of cascades with fixed expected gain $G=8$ under a
      step-function of intensity $\alpha_1=1.2$ and $\alpha_i=1$.  The
      length of each cascade is grown from $n=1,\ldots,10$ with random
      deactivation rates $\beta_i\sim\mathcal{N}((\alpha_1 G)^{-1/n},
      0.05^2)$, and the length of the cascade that achieves maximum
      amplification is recorded. The figure presents the histogram of
      the observed optimal lengths.}
    \label{fig:length-sampled-betas}
\end{figure}

\begin{figure}[tp]
  \centerline{\includegraphics[width=.9\textwidth]{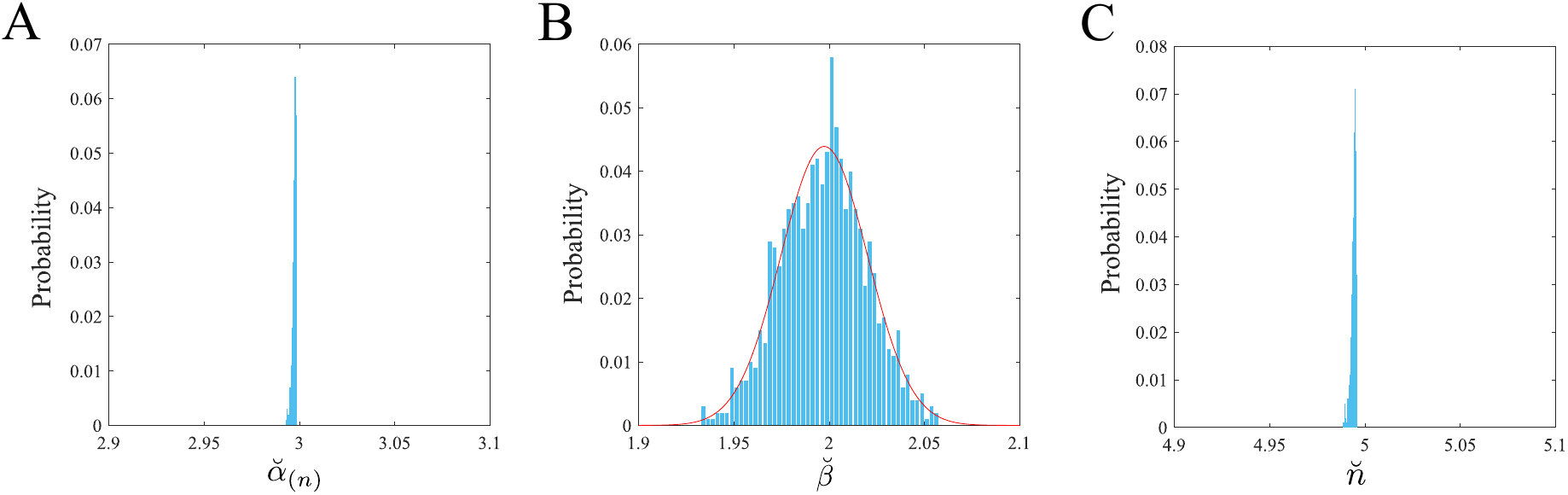}}
  \caption{{\bf Estimation of parameters in random, near-optimal
      cascades.}  Distribution of estimated parameters
    $\breve{\alpha}_{(n)}$, $\breve{\beta}$, and $\breve{n}$ obtained
    when fitting Eq.~\eqref{eq:weak-xn-const-sol} to 1000 cascades in
    which $\alpha_{(n)}=3$, $n=5$, and $\beta_i \sim
    \mathcal{N}\left(2, 0.05^2\right)$ for $i=1,\dots, 5$. The mean of
    the estimated parameters are:
    $\overline{\breve{\alpha}}_{(n)}\approx 2.997$,
    $\overline{\breve{\beta}} \approx1.997$, and $\overline{\breve{n}}
    \approx 4.994$. The red curve shows a fitted normal distribution
    with mean 1.997 and variance $0.022^2$.}
    \label{fig:fits}
\end{figure}

\begin{figure}[tp]
  \centerline{\includegraphics[width=.9\textwidth]{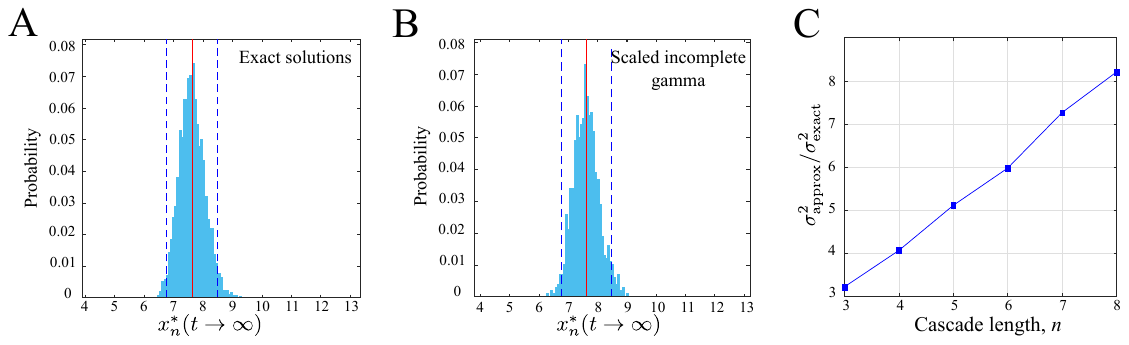}}
    \caption{{\bf Using the analytical approximation for random
        near-optimal cascades.}  We consider a cascade of length $n=5$
      with a step-function input $R=1$, $\alpha_i = 3$ and $\beta_i
      \sim \mathcal{N}\left(2, 0.05^2\right)$.  {\bf A}: Histogram of
      the asymptotic value given by
      Eq.~\eqref{eq:xn-gen-sol-const-stim} using 1000 sampled sets of
      $\{\beta_i\}_{i=1}^5$ (mean=$7.624$, std=$0.043$).  The red,
      continuous line marks the mean of the sample and the dashed
      lines indicate $\pm 2$ standard deviations.  {\bf B}: Histogram
      of the values obtained using the lower incomplete gamma function
      in Eq.~\eqref{eq:weak-xn-const-sol}, with $\beta \sim
      \mathcal{N}\left(2, 0.05^2/n^2 \right)$ (mean=$7.687$,
      std=$0.039$).  {\bf C}: Relationship between the ratio of the
      variances of the full solution and the gamma function
      approximation as a function of $n$.}
    \label{fig:distributions}
\end{figure}

We have tested this idea in
Figures~\ref{fig:length-sampled-betas},~\ref{fig:fits}
and~\ref{fig:distributions}.  
First, we check that cascades with
non-identical deactivation rates still achieve maximal amplification
when the cascade is of finite length, and we characterise the
distribution of cascade lengths observed.
Figure~\ref{fig:length-sampled-betas} shows the histogram of the
cascade length at which maximal amplification is achieved for random
ensembles of cascades.  We consider a step-function input $R(t)=1$
with $\alpha_1=1.2$, and we take as a reference an optimal cascade
with identical activation rates $\alpha_i=1$ for $i>1$ and
deactivation rates $\beta_i = \beta_n = (\alpha_1\,G)^{-1/n}=
9.6^{-1/n}$,  which delivers a gain of $G=8$ with an optimal
finite length of $n=4$~\cite{Chaves2004}.
We then generate 1000 sets of cascades of length $n=1,\dots,10$, with
deactivation rates drawn from a distribution
$\beta_i\sim\mathcal{N}(\bar{\beta}_n , 0.05^2)$,
$\bar{\beta}_n=9.6^{-1/n}$, $i = 1\dots n$ and $n = 1,\dots 10$ and we
record the length at which the maximal amplification occurs. Note that
the mean of the deactivation rates depends on the length of the
cascade. As shown in Fig.~\ref{fig:length-sampled-betas}, near-optimal
cascades (with normally distributed $\beta_i$ with mean $\bar{\beta}_n
= 9.6^{-1/n} $) achieve maximal amplification for lengths between
$n=3$ and $5$ in $60.4\%$ of cases.

To test whether we can use the gamma function to estimate the
parameters of cascades in which the deactivation rates are not
identical, we simulated 1000 cascades under a 
step-function input $R(t)=1$, with $\alpha_{(n)}=3$, $n=5$,
and random deactivation rates $\beta_i \sim \mathcal{N}\left(2,
0.05^2\right)$.  In each cascade we fitted the parameters 
$\breve{\alpha}_{(n)}, \breve{\beta}, \breve{n}$ in
Eq.~\eqref{eq:weak-xn-const-sol} to the `observed' $x_5^*(t)$.
Figure~\ref{fig:fits} shows the histograms of the fitted parameters
for the 1000 random cascades.  The fitted parameters are close to
their `true' values, with the distributions of $\breve{\alpha}_{(n)}$
and $\breve n$ peaked close to their `true' values, and the
distribution of estimated deactivation rates $\breve{\beta}$ normally
distributed around its `true' value.

To check that the outputs for (random) near-optimal cascades can be
well approximated using the gamma function expressions, we show in
Figure~\ref{fig:distributions} that the distribution of asymptotic
values of an ensemble of cascades governed
by~(\ref{eq:xn-gen-sol-const-stim}) with $\beta_i \sim
\mathcal{N}\left(\bar{\beta}_n, \sigma^2\right)$ matches the
distribution obtained from the gamma function
representation~(\ref{eq:weak-xn-const-sol}) with $\beta_i \sim
\mathcal{N}\left(\bar{\beta}_n, \sigma^2/n\right)$.  Hence, the gamma
function form can be used for near-optimal cascades with random
variability in the deactivation parameters, by scaling the variance of
the deactivation rates by the length of the cascade
(Fig.~\ref{fig:distributions}C).

\subsection{Cascade reordering: lumped representation of identical blocks}

As another deviation from strict optimality, we examine how the
output of a weakly-activated cascade is modified when a
single protein in the cascade has a different deactivation rate.  For
instance, Ref.~\cite{Chaves2004} considered an auxiliary protein with
different deactivation rate at the end of the cascade. We study the
effect of such a `perturbation', and the effect of the position of the
perturbation in the cascade.

\begin{figure}[tp]
  \centerline{\includegraphics[width=0.8\textwidth]{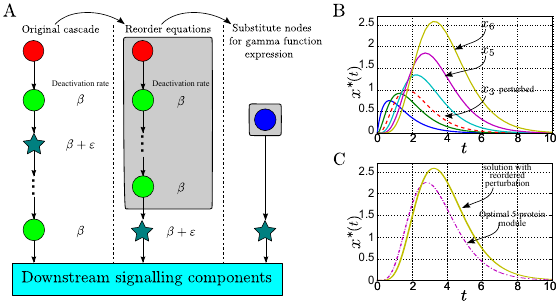}}
    \caption{{\bf Cascade reordering and lumping of
        identical protein blocks into sub-cascades.}   {\bf A}: A linear
      \mbox{$\varepsilon$-perturbed} cascade model;
      the input (red node) can either be constant or decaying; green
      circle nodes are proteins whose deactivation rates are all
      $\beta$; the blue star node has deactivation rate
      $\beta+\varepsilon.$ Downstream of the cascade lie other
      components of the signalling pathway. Reordering of the
      equations moves the perturbed deactivation to the bottom of the
      cascade, with no effect on the output of the cascade.  The first
      three equations in the reordered cascade can then be substituted
      for an incomplete gamma function.  {\bf B}: Time-courses of a
      six-protein cascade where the third protein has an
      $\varepsilon$-perturbed deactivation rate following an
      exponentially decaying input.  {\bf C}: The dot-dashed line shows
      the incomplete gamma function of the re-arranged module of five
      unperturbed proteins given by
      Eq.~\eqref{eq:weak-xn-expdecay-sol}, which does \textit{not}
      correspond to any of the time-courses in \textbf{B}. However,
      the output of the cascade (i.e., the time-course of the
      perturbed protein moved to the bottom of the cascade given by
      Eq.~\eqref{eq:weak-expdecay-pert-sol} and shown as a
      continuous-line) coincides with the output of \textbf{B}. }
    \label{fig:cascade-reordering}
\end{figure}

Consider a cascade of $n$ proteins with activation rates $\alpha_j$
and deactivation rates \mbox{$\beta_j=\beta, \,\, \forall j \neq i $},
and \mbox{$\beta_i = \beta + \varepsilon$} for a given \mbox{node
  $i$.}  First, note that from the Laplace transform of $x_n^*(t)$, it
is clear that the \textit{position in the cascade} of the protein with
distinct deactivation $\beta_i$ does not affect the final output:
\begin{equation}
  \La(x_n^*) = \frac{\alpha_{(n)}^n\La(R)}{(\beta +
    s)^{n-1}(\beta+\varepsilon +s)}.
\end{equation}
This fact allows us to reshuffle the equations of linear cascade models,
grouping the blocks with identical deactivation rates, which can thus 
be lumped upstream in the cascade and replaced with the incomplete
gamma function representation. The equations of the perturbed
proteins can be placed downstream and take the gamma function of the
lumped block as an input.  Such reordering can be used to reduce and
simplify the model of a cascade without altering the dynamics or
timescales (Fig.~\ref{fig:cascade-reordering}A).

More explicitly, suppose we have an \mbox{$\varepsilon$-perturbed}
cascade of $(n+1)$ proteins reordered so that the first $n$ proteins all
have deactivation rate $\beta$ and the \mbox{$(n+1)$-th} protein has
rate $\beta+\varepsilon$. For a step-function input $R(t)=1, \,
t \geq 0 $ we use Eq.~(\ref{eq:weak-xn-const-sol}) to summarise the
first $n$ equations, and the equation for the perturbed $(n+1)$-th
protein becomes then
\begin{equation}
  \ddt{x_{n+1}^*} = \alpha_{n+1} \left(\frac{\alpha_{(n)}}{\beta}\right)^n  
  \mathrm{P}(n, \beta t)
  - (\beta + \varepsilon)x_{n+1}^*.
  \label{eq:weak-conststim-pert-ode}
\end{equation}
This equation can be solved analytically to give:
\begin{equation}
  x_{n+1}^*(t) = \frac{\alpha_{n+1}}{\beta + \varepsilon}
  \left(\frac{\alpha_{(n)}}{\beta}\right)^n \left( 1 - e^{-\beta
    t}\left[\left(\frac{-\beta}{\varepsilon}\right)^n e^{-\varepsilon
      t}
    +\sum_{k=0}^{n-1}\frac{(\varepsilon^{n-k}-(-\beta)^{n-k})(\beta
      t)^k} {\varepsilon^{n-k}k!}\right]\right),
  \label{eq:weak-conststim-pert-sol}
\end{equation}
where we have assumed the initial condition $x_{n+1}^*(0)=0$.

Likewise, when the input is exponentially decaying $R(t)= e^{-\lambda
  t}$, we have that
\begin{equation}
  \La(x_{n+1}^*) =
  \frac{\alpha_{(n)}^n}{(s+\lambda)(s+\beta)^{n-1}(s+\beta+\varepsilon)}.
\end{equation}
When the initial condition is $x_{n+1}^*(0)=0$, the analytical solution
for $\beta\neq \lambda$ is:
\begin{align}
  x_{n+1}^*(t) = \frac{ \alpha_{n+1}}{\beta - \lambda + \varepsilon} &
  \left(\frac{\alpha_{(n)}}{\beta -\lambda}\right)^n \left[e^{-\lambda
      t} + \frac{e^{-(\beta+\varepsilon)t}}{\varepsilon^n} -
    \right. \nonumber \\ & \left. - e^{-\beta t}\sum_{k=0}^{n-1}
    \frac{\left(\varepsilon^{n-k} - (\lambda -
      \beta)^{n-k}\right)(\beta-\lambda)^k t^k} {\varepsilon^{n-k} k!}
    \right].
  \label{eq:weak-expdecay-pert-sol}
\end{align}
When $\lambda=\beta$ the solution is 
\begin{equation}
   x_{n+1}^*(t) =
   \left(\frac{\alpha_{(n+1)}}{\varepsilon}\right)^{n+1}e^{-\beta
     t}\left[
     \varepsilon^n\sum_{k=0}^n\frac{(-1)^kt^{n-k}}{\varepsilon^k
       (n-k)!}  + (-1)^{n+1}e^{-\varepsilon t}\right].
   \label{eq:weak-expdecay-pert-sol-2}
\end{equation}  

We illustrate these points in Figure~\ref{fig:cascade-reordering}.
Figure~\ref{fig:cascade-reordering}B shows the time course of a
cascade with six proteins in which the deactivation of the third
protein is perturbed. We then reorder the equations so that the
perturbed one lies at the bottom. Figure~\ref{fig:cascade-reordering}C
shows the output of the first 5 reordered equations, given by the
gamma function expression~\eqref{eq:weak-xn-expdecay-sol})(dot-dashed
line), and the analytical solution of the perturbed protein (which is
now the output of the cascade, continuous line), given by
Eq.~\eqref{eq:weak-expdecay-pert-sol}. Note how the time-courses of
the fifth protein in the original and rearranged cascades are
different, yet the time course of the sixth protein is identical in
both cases, as per our solution. Given the results for random cascades
presented above, this approach can be applied to lump sub-cascades of
proteins with similar deactivation rates which can then be described
compactly through their corresponding gamma function modules.

\subsection{Simplified modules for activation cascades and delay-differential equation models}

Experimental observations in signalling cascades are typically
concerned with the amplification, distortion and delay introduced in the output.  
As discussed above, when using ODE models, 
delays are usually incorporated through the addition
of extra equations (and their corresponding extra variables and parameters) 
corresponding to unmeasured, hidden components, steps, or processes in the
cascade~\cite{Stark2007}.  This approach can lead to large
(high-dimensional) models with many unobservable variables and high
numbers of parameters to be identified or fitted~\cite{Bar-Or2000,Hoefer2002}.
Alternatively, modellers often use delay differential equations (DDEs)
\nomenclature{DDE}{Delay differential equation} to account for the lag
between an event and its effect~\cite{Bernard2006, Colijn2005,
  Monk2003}.  In a DDE, the activity of a variable depends on the
state of the system a time $\tau$ in the past:
$$\ddt{\mbf{x^*}} = \mbf{f}(\mbf{x^*}(t-\tau)),$$ where the parameter
$\tau \geq 0$ is the delay. Although linear systems of DDEs 
can in principle be solved analytically using
infinite series involving the Lambert function~\cite{Bellman1963,
  Yi2006}, such solutions are often impractical to use.

\begin{figure}[tp]
  \centerline{\includegraphics[width=0.8\textwidth]{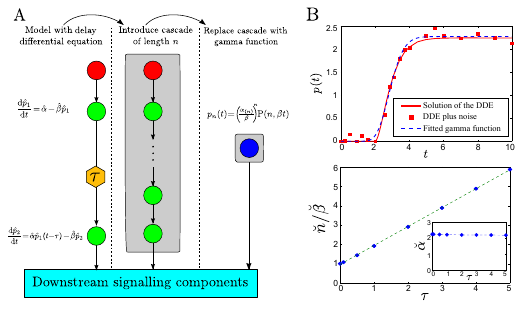}}
    \caption{{\bf Using linear activation cascades to
        replace delay differential equations.}  {\bf A:} An input
      signal (red node) activates a node in a signalling pathway. The
      bottom node responds with a delay $\tau$.  The delay in the
      equation can be substituted with a linear cascade of unknown
      length $n$, which in turn can be described by a lower incomplete
      gamma function.  {\bf B}: Top: The continuous line is the full
      solution to the DDE~(\ref{eq:delay-de}) and the squares are
      `data points' taken from this solution with added random noise.
      The dashed line is the fit of the data using a lower
      incomplete gamma function. Bottom: The ratio of the fitted
      $\breve{n}/\breve{\beta}$ scales linearly with $\tau$, the delay in the original data: 
      \mbox{$\breve{n}/\breve{\beta} = 0.962 + 0.977\tau$} (dashed-line).
      Inset: The fitted parameter $\breve{\alpha}$ remains constant with varying $\tau$.}
    \label{fig:delayReplace}
\end{figure}

We have checked that our results can be applied to model
simple delays in linear activation cascades, leading to concise ODE models that
capture the delay through the gamma function terms
without the need to rely on DDEs (Fig.~\ref{fig:delayReplace}A).
As an example, consider a system with delay modelled
with the linear~DDE:
\begin{align}
  \ddt{\hat{p}_1} & = \hat{\alpha} - \hat{\beta} \, \hat{p}_1,
  \nonumber \\ \ddt{\hat{p}_2} &= \hat{\alpha} \, \hat{p}_1(t-\tau)
    - \hat{\beta} \, \hat{p}_2.
    \label{eq:delay-de}
\end{align}
Figure~\ref{fig:delayReplace}B (top) shows the simulated time course
of $\hat{p}_2(t)$ (continuous line) when $\hat \alpha=2$, $\hat
\beta=3$, and $\tau=2$ with initial conditions
$\hat{p}_1(0)=\hat{p}_2(0)=0$. This series was numerically obtained
with the {\tt dde23} solver in Matlab.  We then generate our `observed
data' by sampling $\hat{p}_2$ at various time points and adding
observational random noise from a distribution $\mathcal{N}(0,
0.05^2)$.

We then fit this noisy data to our gamma function
expression~\eqref{eq:weak-xn-const-sol}:
\begin{equation}
  p_n(t) = \left(\frac{\alpha_{(n)}}{\beta}\right)^n\mathrm{P}(n, \beta t)
  \approx \hat{p}_2(t),
  \label{eq:delay-approx}
\end{equation}
and we estimate the corresponding parameters.
Figure~\ref{fig:delayReplace}B (top) shows the fit, as obtained with
the Squeeze-and-breathe algorithm~\cite{Beguerisse2012b}, with
estimated parameters $\breve{\alpha}_{(n)} \approx 2.270$,
$\breve{\beta} \approx 7.530$, and $ \breve{n} \approx 22.107$.

To explore the connection between the parameters of the DDE and the
best fit activation cascade model, we simulate the
DDE~\eqref{eq:delay-de} with parameters $\hat{\alpha}=2$ and $\hat
\beta=3$ for different values of the delay $\tau \in [0,5]$ and fit
models as above.  The dependence of the fitted parameters and $\tau$
is shown in Fig.~\ref{fig:delayReplace}B (bottom plot). Reassuringly,
the amplification factor $\breve{\alpha}_{(n)}$ 
  remains relatively constant, whereas
the ratio $\breve{n}/\breve{\beta}$ grows linearly with $\tau$. This can be expected
from the simple argument that the time delay $\tau$ in the DDE
should be related to the accumulated time needed to traverse $n$ sequential
steps with duration $1/\beta$. 
Hence, a DDE with delay $\tau$ can be approximated with a linear cascade, 
by tuning \textit{both} the length and the deactivation rate of the cascade, 
i.e.,  $\tau \sim n/\beta -1$.

In Ref.~\cite{Beguerisse2012} we have used the approach described here
to introduce delays in the antioxidant responses of guard cells to
abscisic acid and ethylene stimuli during stomatal closure in an ODE
model.

\section{Discussion}

In this work, the classic model of activation cascades in the weakly
activated regime~\cite{Heinrich2002} has been re-examined.  We have
considered the important case where all deactivation rates of the
components of the cascade are identical, which was shown to
provide optimal amplification in Ref.~\cite{Chaves2004}. Our results
show that the output of optimal cascades can be represented exactly by
lower incomplete gamma functions, and we show numerically that even
when the cascades are near optimal (i.e., when the deactivation rates are
iid normal random variables) a gamma function can
summarise the cascade by an appropriate rescaling of the parameters.
We also show that the position of a protein in the cascade does not
affect the final output, so that blocks of proteins with identical
deactivation rates can be lumped and represented with incomplete gamma
functions.  These results allow the reduction of the number of
equations and parameters in ODE models without affecting the dynamics
or the timescales of the system. We have also shown that in some cases
incomplete gamma functions can be used to model delays within
systems of ODEs, as an alternative to delay differential equations.

Beyond its application to enzymatic activation cascades, similar
mathematical models of cascades could be helpful for the
parametrisation and modelling of multi-step transcriptional processes,
an area of active research in Systems and Synthetic
Biology~\cite{Hooshangi2005, Lucius2003, Stricker2008, Wang2011}.
In general, model reduction of systems of differential equations remains a 
challenging and active area of research~\cite{Conzelman2004, prajna2005, Siahaan2008}.
Some methods reduce network models (or modules) based on the topology,
effectively finding a minimal kernel that preserves some aspects of
the dynamics~\cite{Kim2011}. Yet, by only considering the topology of
the system such methods cannot be guaranteed to preserve timescales or
behaviour~\cite{Ingram2006}, and are best suited for boolean
models. As Ref.~\cite{Beguerisse2012} shows, timescales and transients
can be crucially linked to the behaviour of a model and cannot be
ignored in many cases. Our work introduces a simplified, compact
description that can serve to consider delays in ODE models for
Systems and Synthetic Biology, and to fit data from experimental
observations.


\section*{Acknowledgments}
MBD acknowledges support from the James S. McDonnell Foundation
Postdoctoral Program in Complexity Science through a Complex Systems
Fellowship Award (\#220020349-CS/PD Fellow) and a BBSRC-Microsoft
Research Dorothy Hodgkin Postgraduate Award. MB acknowledges funding
from the EPSRC through grants EP/I017267/1 and EP/N014529/1, and from
BBSRC through grant BB/G020434/1.  The authors thank Piers J.\ Ingram
and Lars Bergemann for many useful conversations.
%
No new data was collected in the course of this research.

{\footnotesize
\bibliographystyle{siam}
\bibliography{bib-cascades}
}

\end{document}